\documentclass[twocolumn,11pt]{article}
\usepackage{times}
\usepackage{graphicx}
\usepackage{algorithm}% http://ctan.org/pkg/algorithms
\usepackage{algpseudocode}% http://ctan.org/pkg/algorithmicx
\usepackage{enumitem}
\usepackage{setspace}
\usepackage{mathtools}
\usepackage{amsmath}
\setlist{nolistsep}
\begin{document}

\global\def\refname{{\normalsize \it References:}}
\onehalfspacing
\baselineskip 12.5pt
%
%
% TITLE, AUTHOR, ABSTRACT, KEYWORDS
%
\title{\LARGE \bf Active-Threaded Algorithms for Provenance Cognition in the Cloud preserving Low Overhead and Fault Tolerance }

\date{}

\author{\hspace*{-10pt}
\begin{minipage}[t]{2.7in} \normalsize \baselineskip 12.5pt
\centerline{ASIF IMRAN, EMON KUMAR DEY, KAZI SAKIB}
\centerline{Institute of Information Technology}
\centerline{University of Dhaka, Bangladesh}
\centerline{asif.imran.anik@gmail.com, emonkd@iit.du.ac.bd, sakib@univdhaka.edu}
\end{minipage} \kern 0in
\\ \\
\begin{minipage}[t]{2.7in} \normalsize \baselineskip 12.5pt
\centerline{M. ABDULLAH-AL-WADUD}
\centerline{Department of Industrial and Management Engineering}
\centerline{Hankuk University of Foreign Studies, South Korea}
\centerline{wadud@hufs.ac.kr}
\end{minipage}\kern 0in
%
% If you are three authors then you can use three mini--pages
% instead of two. Their horizontal size must be less than 2.7in
% indicated above. It can be e.g. 2.3in. However, you must pay
% attention that you do not exceed the total width of the text.
%
\\ \\ \hspace*{-10pt}
\begin{minipage}[b]{6.9in} \normalsize
\baselineskip 12.5pt {\it Abstract:}
% The text of the abstract follows.
Provenance is the derivation history of information about the origin of data and processes. For a highly dynamic system such as the cloud, provenance must be effectively detected to be used as proves to ensure accountability during digital forensic investigations. This paper proposes active-threaded provenance cognition algorithms that ensure effective and high speed detection of provenance information in the activity layer of the cloud. The algorithms also support encapsulation of the provenance information on specific targets. Performance evaluation of the proposed algorithms reveal mean delay of 8.198 seconds that is below the pre-defined benchmark of 10 seconds. Standard deviation and cumulative frequencies for delays are found to be 1.434 and 45.1\% respectively.
\\ [4mm] {\it Key--Words:}
% The key-words follow.
Provenance Detection, Cloud Computing, Provenance Encapsulation, Active-Threading
\end{minipage}
\vspace{-10pt}}

\maketitle

\thispagestyle{empty} \pagestyle{empty}
% numbers of pages are supplemented by the editor
%
% THE BEGINNING OF THE TEXT
%
\section{Introduction}
\label{S1} \vspace{-4pt}

 Provenance is the meta-data that represents information about the operations executed on specific processes at the activity layer. Cloud Computing is the dynamic provisioning of resources from a shared resource pool that stores critical information of the customers and requires provenance of those to achieve accountability. Real life service providers execute mission critical processes and store high-value information in the cloud \cite{pohly2012hi}. Hence provenance needs to be detected for tracking the processes running at the activity layer of the cloud to ensure monitoring of the stored data and aid in digital forensics.%Cloud is an evolving technology, the basis of which is virtualization of physical servers to form virtual machines (vm) instances \cite{armbrust2010view}. This paper studies existing provenance schemes for the cloud and identifies the major challenges involved in cloud provenance. The use of provenance to ensure, security, privacy, audability and accountability of cloud has been proposed.

Capturing provenance and encapsulating those on specific processes in a dynamic system such as the cloud involves significant research challenges, since a large number of dynamic processes are involved. Traditional provenance cognition schemes for distributed systems consist of frameworks capturing system log files that has high overheads in terms of execution time. However provenance detection in the cloud must be real time and dynamic, since a large number of physical and virtual machine (vm) instances are involved. Algorithms that capture and bind provenance information to the original data files in real time without causing significant time overhead and loss of fault tolerance have not been considered for the cloud in earlier studies. More precisely, the following research issues need to be addressed.

\begin{enumerate}
\item Novel algorithms with active-threading capability that capture provenance in the cloud and encapsulate the data file with the provenance information without causing significant overheads.
\item Empirical investigation to analyze the performance and overheads of the proposed active-threading algorithms at the activity layer and comparing the obtained results with benchmarks.
\end{enumerate}

Existing mechanisms are not suitable for cloud provenance since they cause significant overhead in terms of intermessage transmission time and cause the cloud processes to slow down at the activity layer. \emph{Ko. et al} identified accountability and audability as prime challenges for cloud computing \cite{armbrust2010view}. Through real life scenarios of malicious cloud attacks, the vulnerability of cloud to attacks was shown. However, the research did not focus on how provenance detection can effectively prevent such attacks. Technical and procedural approaches to ensure security of cloud were discussed in \cite{armbrust2010view}. Schemes such as identification of anonymous login and anonymous users through pseudomonas file read were proposed in \cite{asif2013cloudniagara} without performance comparison with standard benchmarks.

Based on the research issues identified above, this paper proposes two novel algorithms called ProvCapsule and ProOCal to detect provenance at low overhead in the activity layer. The activity layer deals with all the user inputs and triggers in the cloud. The provenance encapsulation algorithms achieves the functionality without causing significant delay in transmission time by using active-threading modules. The algorithms capture provenance by treating data files as objects and identifies critical information such as access time, number of operations, type of operations, the executing entity, senders and receivers of the data under consideration. The tracked provenance are then stored in separate meta files that are bounded to the original data file. This binding is based on identity number (id-number) allocation to both the provenance and original files and it helps forensic experts track provenance information to specific targets.

The performance of the proposed algorithms have been analyzed in the environment of real life service providers. Provenance at the activity layers of 936 vm-instances were tracked for files ranging from 512-3072 MegaBytes (MB). Next the provenance files encapsulated the original files and out of 936 cases, the average mean time was found to be 8.198 seconds. Standard deviation was determined to be 1.434 which is desirable with an estimated variance 2.056 considering the large number of vm-instances that were used. The results show desirable performance as the overhead incurred in terms of delay because of the encapsulation module of the algorithm is lower than 10 seconds. Cumulative frequencies were obtained to be 45.1\% of the files had about 7-8 seconds transmission and global increase in time for provenance, showing the effectiveness of the algorithms.

\section{Related Work}
\label{S2} \vspace{-4pt}

The remote storage of data and remote computation is considered as a critical risk for both the cloud service provider and customer \cite{storage2010case}. Accountability must be ensured for both customers and service providers of the cloud for ensuring security \cite{storage2010case}, \cite{zissis2012addressing}. In case of a problem, both parties must be made aware about the liabilities and consequences of the issue. Secure auditing, recording and evidence of information stored on the cloud was proposed as probable solutions to the issue. However, the importance of log based analysis of provenance data has been covered to a little extent.

Accountability of cloud services was identified as a complex challenge to achieve by the cloud developers \cite{ko2011trustcloud}. Through the description of real life scenarios which included storing images on cloud servers, the real time tracing of operations executed on the cloud were identified. At the same time file-life-cycle logging, file-change logging was identified to be primary challenges \cite{ko2011trustcloud}.  It was described how malicious attackers try to transfer files on the cloud to servers outside using email services. Regulations which hinder the blocking of such transfer were also identified. The study does not provide a provenance detection scheme that identified both file-centric and system-centric operation on the cloud and can be successfully used to identify and block such attacks.

Technical and procedural approaches to ensure privacy is a key goal for software engineers when building a cloud for the production environment \cite{haeberlen2010case}. Lower levels of privacy through the use of privacy management tools for system level operations was focused by the authors. Use of pseudonymisation tools to hide the name of real users is proposed. These technologies include anonymous login, pseudonymous identity number and email addresses. Cryptographic mechanisms are used to ensure that the security and integrity of login information. However the use of log-based provenance to aid in data-forensics is discussed to a little extent.

Privacy aware provenance of scientific workflows is stated in terms of data module and privacy policy \cite{mowbray2012enhancing}, \cite{lu2010secure}. The authors raised questions regarding provision of unlimited or fixed number of allowable provenance queries. The provenance information is to be made available to the user assuring the security and privacy of the information. Composite and simple workflows of provenance data were represented as acyclic graphs and then inferences were obtained from those. However, necessity of user defined provenance information and ensuring management of the captured provenance were not considered.

Recent technologies for provenance detection and scientific workflow systems have been identified in \cite{abbadi2012framework}. The data are stored in systems using specialized Semantic Web Languages and XML dialects that are stored as files and the tuples stored in relational databases. Workflow processes contained input of the experiments and the output produced by processing the data. The study is only concerned with scientific workflow systems and does not consider provenance detection, storage and management of system-level files or processes involved in the activity layer.

The above discussion amplifies the importance of novel algorithms that can effectively detect provenance and cause low overhead in terms of data transmission time. The importance of parallel active-threading algorithms to preserve the fault tolerance capability of the cloud and encapsulate provenance on the data files is manifold. The need for performance analysis and comparison with defined benchmarks is also important for the effectiveness of the algorithms.

\section{Proposed Algorithms for Provenance Cognition at the Activity Layer}
\label{S6} \vspace{-4pt}

This section proposes algorithms that are necessary to capture the provenance and store it in a provenance file. Next the file is encapsulated as a metafile together with the original data file. The algorithm treats every data file as individual objects and detects provenance for those.

Active-Threaded provenance cognition algorithms ensure that the provenance is detected without having an affect on  the fault tolerance ability of the cloud environment through rapid encapsulation. The next subsection describes ProvCapsule, the novel provenance detection and encapsulation algorithm proposed here.

\subsection{ProvCapsule: Algorithm that Captures and Binds Provenance to Cloud Data}
\vspace{-4pt}

Initially, each file contains a provenance metafile called $F_{PROV}$ identified by a unique id $B_{ID}$. Original files identified by $F_{ORG}$ are stored in $DataFile[x]$ array. $F_{PROV}$ are stored in $ProvFile[y]$ array. Each $F_{PROV}$ encapsulates specific $F_{ORG}$ together with its provenance information. Active-Threading is ensured as the algorithms read each $F_{ORG}$ and treat those as individual objects. Next the methodology sets $B_{ID}$ for each $F_{ORG}$ and $F_{PROV}$ to identify those uniquely.

\begin{algorithm}
 \begin{algorithmic}[1]
 \Procedure{ProvCapsule}{$a,b$}
 \State $B_{ID}\gets key$ and $F_{ORG}\gets DataFile[x]$ and
 \State $F_{PROV}\gets ProvFile[y]$ and
 \State $P_{SB}\gets Loc[z]$
 \While{$B_{ID}\not=0$}
 \State Read inputs in $F_{ORG}$
 \State Record for every object $a$ in $DataFile[x]$
 \While {$a$ = $DataFile[x]$}
 \State $B_{ID}=a.B_{ID}$ and
 \State $F_{ORG}$=$F_{ORG+1}$
 \State $DataFile[x+1]$=$F_{ORG}$
 \State continue
 \EndWhile
 \State Record for every object $c$ in $ProvFile[y]$
 \State If $B_{ID}.F_{PROV}$=$c$.$F_{PROV}$
 \State $B_{ID}.F_{PROV}\gets B_{ID}.F_{ORG}$
 \State $B_{ID}\gets B_{ID}+1$
 \EndWhile
 \State $ProvFile[y]$=($Loc[P_{SB}]$,$P_{SB}.Exec$,$B_{ID}$)
 \State \textbf{return} $B_{ID}.F_{PROV}$ \& $length$.$DataFile[x]$
  \EndProcedure
 \end{algorithmic}
 \caption{Provenance Capsule at Cloud Activity Layer}
 \label{euclid}
\end{algorithm}

Once the $B_{ID}$ of an object currently in queue matches an object stored in the $DataFile[x]$ array, it will read the input information $I_{a}$ that is a subset of the original set of inputs $I$. This process is used to detect and associate provenance information such as date, time, operator and operation of specific $DataFile[i]$ into the $ProvFile[j]$. The data are stored in a separate $F_{PROV}$ file which is then bounded to the original file as a metafile. The $F_{ORG}$ is then transmitted in the cloud due to commands of the activity layer.

\subsection{ProvOCal: Algorithm to Map Provenance Information of Memory and Disk Activities}
\vspace{-4pt}

The algorithm for mapping provenance for memory reads and disk writes are discussed here. Two variables $g$ and $h$ are used to compare length of memory read and length of disk writes respectively. After those are recorded, g and h compared with the obtained results. $MInf$ is the variable for storing memory information of $F_{ORG}$ and recording its memory length. $MSum$ adds the length of disjoint specific blocks to calculate the total size of memory required. The result is added to the minimum between one less than the total length or $g$.

\begin{algorithm}
 \begin{algorithmic}[1]
 \Procedure{ProvOCal}{$a,b$}
 \State $g\gets 0$ and $h\gets 0$
 \State $MInf=Loc[P_{SB}+1]$
 \While {$g < len.MInf$}
 \State $MSum=MInf$
 \State $S=min(g,len.MInf-1)$
 \State $MSum=MSum+S$
 \EndWhile
 \While {$h < len.SBInfo$}
 \State $SBSum=SBInf$
 \State $T=min(h,len.SBInf-1)$
 \State $SBSum=SBSum+T$
 \EndWhile
 \State If $len.B_{ID}<len.MemInf$, then
 \State $f \gets len.MemInf - len.B_{ID}$
 \State h = h+1;
 \State \textbf{return} Memory and Disk block consumption for provenance capsule
  \EndProcedure
 \end{algorithmic}
 \caption{ProvOCal Algorithm for Provenance Detection in Memory and Disks}
 \label{euclid}
\end{algorithm}

The calculation of disk location and length of the $DataFile[x]$ and $ProvFile[y]$ is similar to the $MemInf$ calculation, with $SBSum$ saving the final value. Finally the difference in the length of $MSum$ and $SBSum$ are calculated on the basis of the matched $B_{ID}$ to determine whether the file is loaded into memory from disk in a compressed format. Hence effective information from the activity layers in terms of memory and disk information can be identified from the algorithm. Next the value of $h$ is incremented as the algorithm returns the disk and memory consumption of the $F_{ORG}$ object at the activity layer of memory management.

\section{System Design for ProvCapsule and ProvOCal}
\label{S4} \vspace{-4pt}

\begin{figure}[ht!]
\centering
\includegraphics[width=90mm]{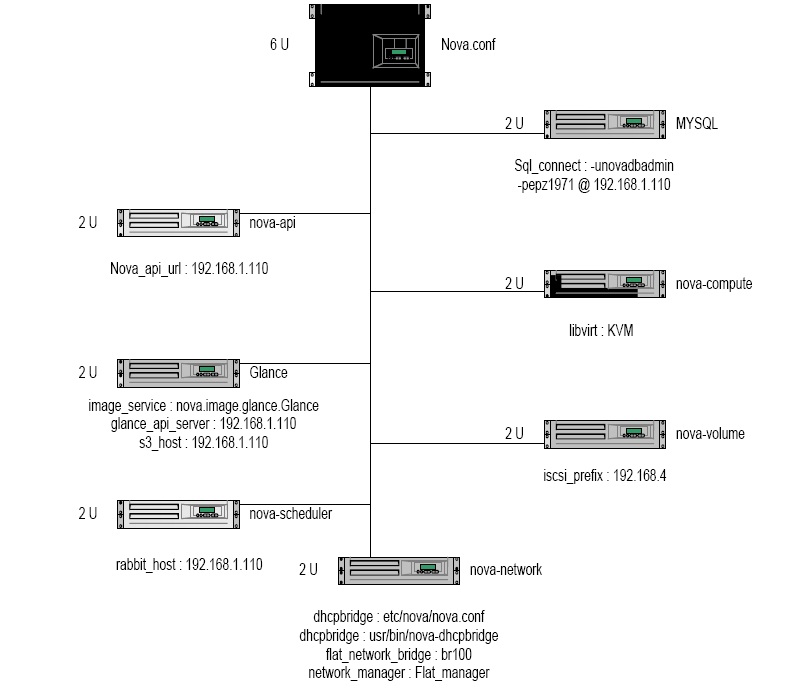}
\caption{Cloud Architecture capable of Provenance Detection}
\label{overflow}
\end{figure}

The proposed algorithms are implemented in real life environment of Commercial Cloud Service provider based on OpenStack. The cloud infrastructure has 10 physical nodes and a controller. The algorithms are implemented on the controller node of the cloud. Hence the resource nodes are bound to provide provenance information as directed by the master cloud controller. The processes include operations on files of various sizes ranging from 515 to 3072 Megabytes (MB)used by 960 vm-instances used by the clients of the cloud service provider. The ProvCapsule and ProvOCal algorithms were implemented in the cloud master controller server.

The vm-instances to which provenance was encapsulated belonged to customers of the cloud provider who ran Business Apps such as inventory management software as Software as a Service (SaaS). Hence the algorithms were tested on real life cloud services.

OpenStack Grizzly Cloud on Ubuntu 12.04 Long Term Service (LTS) server formed the experimental environment. The experiments in the case study included running the applications in virtual machine instances of the cloud. Each instance is allocated 2 GigaBytes of Random Access Memory (RAM) and 2 Central Processing Unit (CPU) cores. Next, the resource consumption and delay of the processes due to encapsulation of the provenance data using the proposed algorithms are identified.

\subsection{Sample Scenarios subject to Provenance Detection}
\vspace{-4pt}

In an operating system each event is divided into a series of atomic steps. Each of the steps can be derived from the atomic actions and the kernel system calls. A specific pattern must be followed \cite{akoush2013hadoopprov}, \cite{slipetskyy2011security}. This pattern is called the signature of the operation which is necessary for provenance detection since it characterizes and provides behavioral information of the activity of that process.

Analyzing and identifying signatures of text based file operations is a prime goal of provenance detection of file activities in cloud computing environments. Signatures of text oriented file system calls of Linux Operating Systems are listed below.

\subsection{Creation of a File}
\vspace{-4pt}

\begin{itemize}
\item Issue command to \textbf{CREATE} text file in a pre-specified process and directory \textbf{AND}
\end{itemize}

\begin{figure}[ht!]
\centering
\includegraphics[width=60mm]{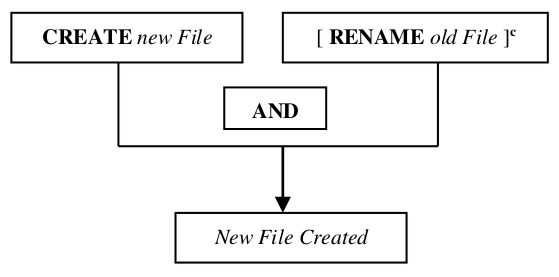}
\caption{Provenance Capture of File Creation and Delete}
\label{overflow}
\end{figure}

\begin{itemize}
\item The \textbf{RENAME} operation is \textbf{NOT} executed on the same File in the same process, i.e, [\textbf{RENAME} old File]
\end{itemize}

\subsection{Copying a File or its contents to another File}
\vspace{-4pt}

\begin{itemize}
\item Issue \textbf{COPY} command to the system call, \textbf{OR}
\end{itemize}

\begin{itemize}
\item \textbf{CREATE} a new File in the same or new directory and \textbf{COPY} the contents of the old File and paste them on the new File.
\end{itemize}

At the Data Layer, the main objective is to analyze the logs which are collected at the System levels \cite{gias2013ivridio}. Log files of Cloud Computing provide information that can be used to collect end-to-end system provenance. The provenance information collected in this way will be highly useful to achieve security and trust on behalf of the cloud customers from the cloud service providers.
\begin{figure}[ht!]
\centering
\includegraphics[width=60mm]{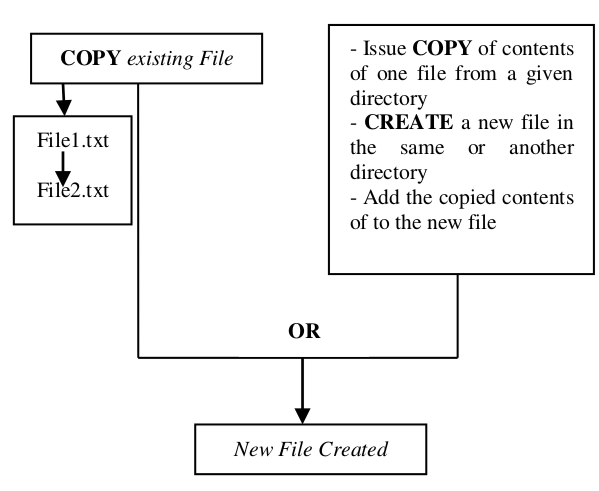}
\caption{File Copy Operation at the Activity Layer}
\label{overflow}
\end{figure}

Based on the analysis of the cloud computing infrastructure and existing provenance models, provenance detection algorithms must be placed forward with an aim of detecting file operations which have possibility of enabling data leakage. ProvCapsule and ProvOCal are such algorithms that aim to drive real time provenance from cloud  activity layer without hampering the fault tolerance of the cloud.

The overhead incurred for encapsulating provenance in terms of global delay and message transmission delay needs to be measured for the algorithms proposed here. The results of the algorithms must be compared to established benchmarks to ensure that the overhead incurred does not exceed the pre-defined limit \cite{slipetskyy2011security}. The following sections identifies the performance of the algorithms in terms of overhead incurred for transmission delay and global delay for encapsulating provenance.

\section{Analysis of Results}
\begin{table}
\caption{Global Delay, Inter-Message Delay (IMT) and Retries of provenance capture}
\begin{tabular}{ l||*{6}{c}r}
\hline
VM-Count & Size & TGD(s) & IMT(s) & Retries\\
\hline
140 & 512 & 817 & 28.9 & 0  \\
100 & 1024 & 867 & 91.4 & 0 \\
160 & 1536 & 892 & 90.9 & 0  \\
162 & 2048 & 948 & 102.7 & 7  \\
184 & 2560 & 1072 & 169.3 & 3  \\
190 & 3072 & 1118 & 281.61 & 7 \\
\hline
\end{tabular}
\end{table}

The results of overhead calculation for the proposed algorithms are tabulated in Table 1 and Table 2. The number of vm-counts that transferred files of specific sizes ranging from 512-3072 MB are shown in column $VM-Count$. The Total Global Delay $TGD$ is shown for the specific vm-instance counts in seconds. In addition the inter-message transmission delay $IMT$ are also shown in seconds together with the number of retries of provenance encapsulation.

The algorithms were implemented in the cloud controller that hosted 936 vm-instances. Hence there were a finite population of vm-instances $X_{i}$ such that $X_{i}$ = \begin{math}
  \begin{pmatrix}
    x_{1},&x_{2},&\cdots&,x_{n}
  \end{pmatrix}
\end{math}. There are a finite population of vm-instances so the mean delay incurred by all those is finite as well. Hence we detect mean, variance and standard deviation for any $X_{i}$ as,

\begin{equation}\overline{X_{i}} = \frac{\sum_{i=1}^{n} X_{i}}{n}\end{equation}

The value can be obtained for a large number of observations n. The variance of the delays in different ranges of Global Delays ($TBD$) and Inter-Message Time ($IMT$) is given by,

\begin{equation}{S^2(n)} = \frac{\sum_{i=1}^{n} (X_{i}-\overline{X(n)})^2}{n-1}\end{equation}

The closeness of the variance \begin{math}S^2_{n}\end{math} to mean \begin{math}\mu\end{math} can be determined as \begin{math}Var(\overline{X(n)})\end{math} is the ratio of total variance and number of occurrences for a given period $t_{i}$.

\begin{equation}{STD.dev} = \sqrt\frac{\sum_{i=1}^{n} (X_{i}-\overline{X(n)})^2}{n(n-1)}\end{equation}
\begin{figure}[ht!]
\centering
\includegraphics[width=3.76in,natwidth=210,natheight=942]{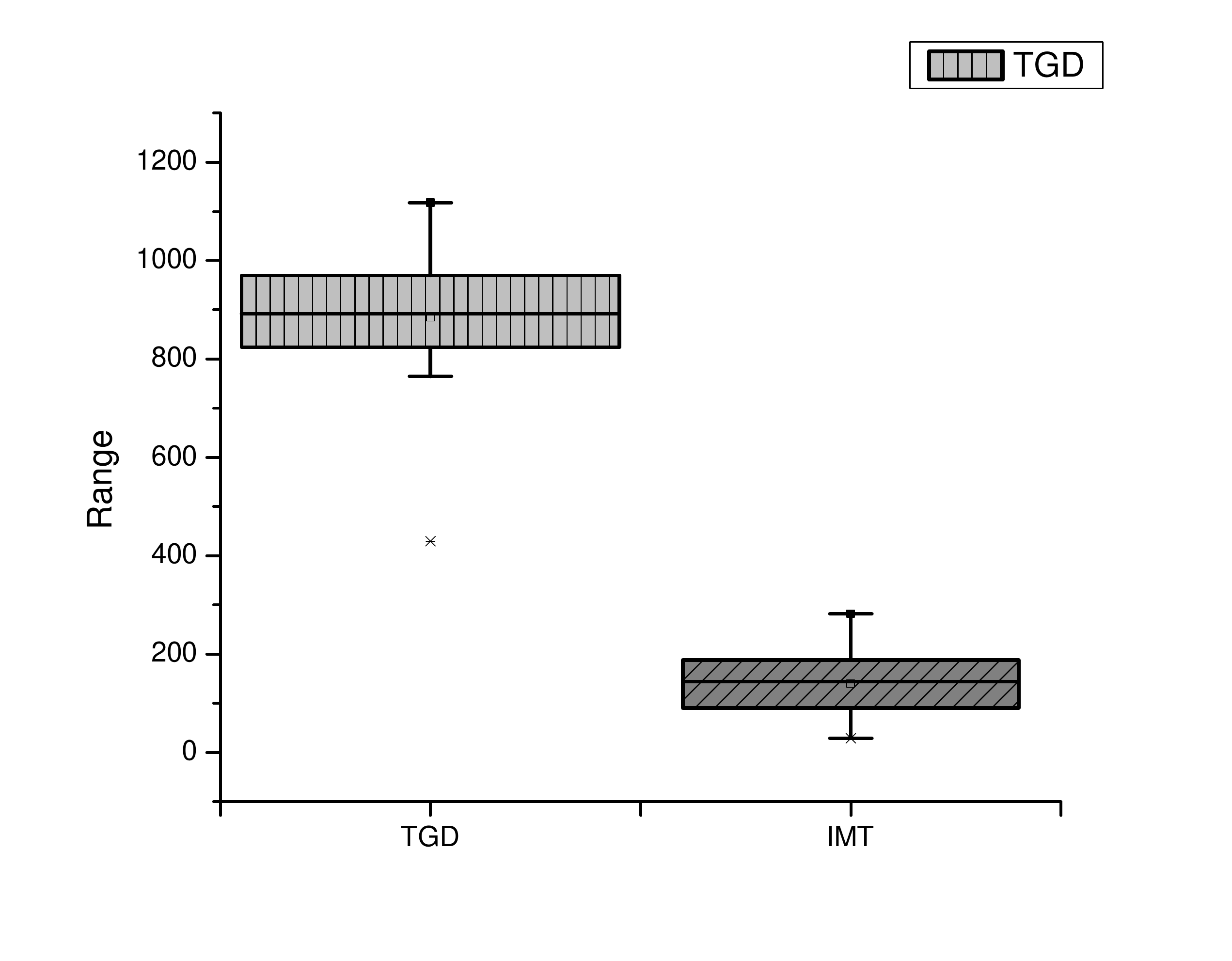}
\caption{Global Delays and Inter-message Delay Distributions}
\label{result1}
\end{figure}

\begin{figure}[ht!]
\centering
\includegraphics[width=70mm]{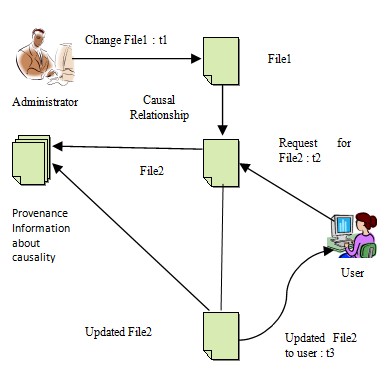}
\caption{Management of Provenance Information in the Cloud}
\label{overflow}
\end{figure}

\begin{figure}[ht!]
\centering
\includegraphics[width=70mm]{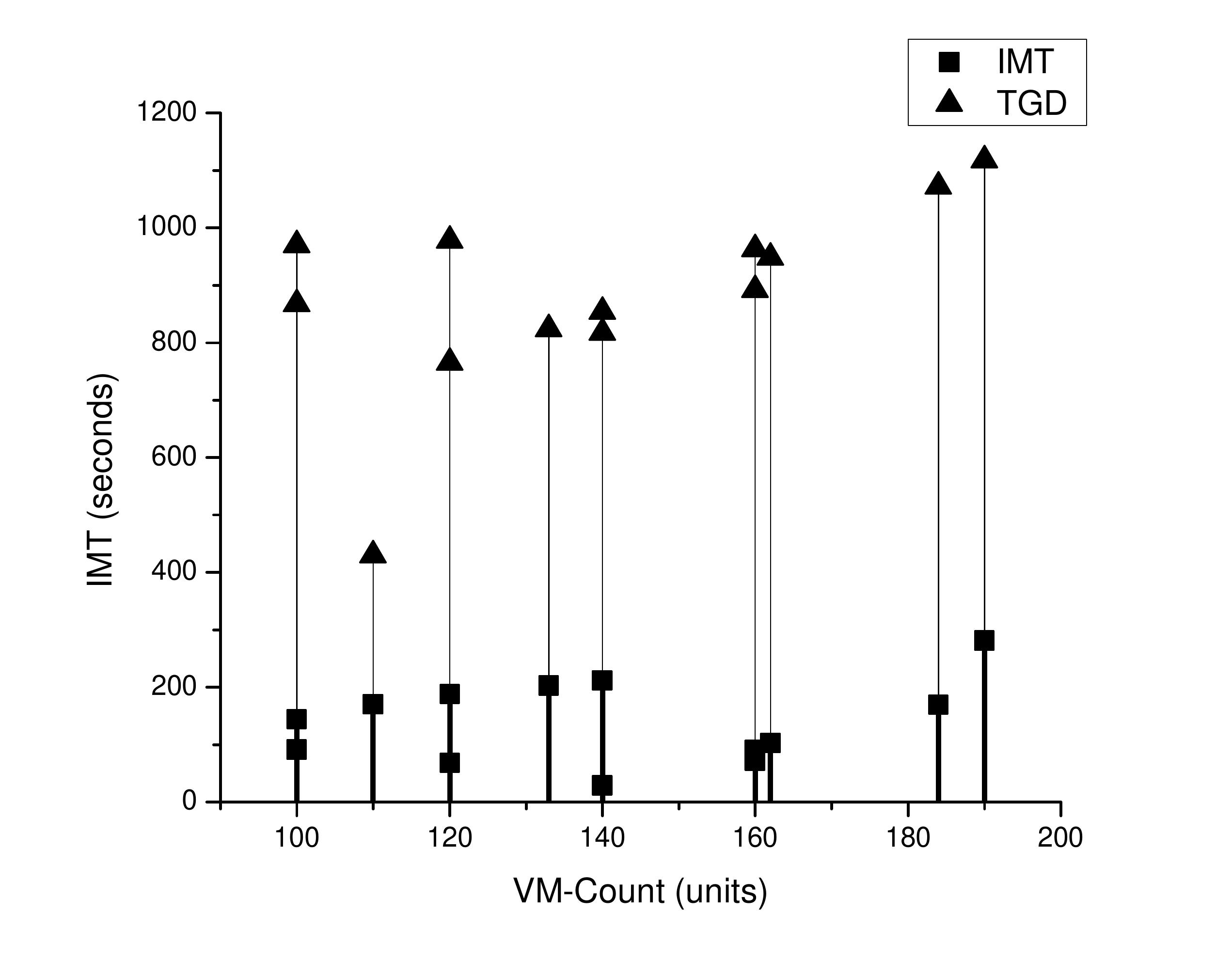}
\caption{TGD and IMD counts}
\label{overflow}
\end{figure}
 \begin{table}
\caption{Mean Delay and Standard Deviation of Provenance Encapsulation}
\begin{tabular}{ l||*{6}{c}r}
\hline
Size & M.Cont. & M.Time(s) & M.Delay(s) & STD.Dev\\
\hline
512 & 5.104 & 6.042 &  &   \\
100 & 5.924 & 8.584 &  &  \\
160 & 6.368 & 7.505 & 8.198 & 1.434  \\
162 & 7.011 & 8.866 &  &   \\
184 & 8.026 & 9.990 &  &  \\
\hline
\end{tabular}
\end{table}
\begin{table}
\caption{Frequencies and Cumulative Frequencies of different Delay Ranges}
\begin{tabular}{ l||*{6}{c}r}
\hline
Delay Range & Frequency & Cumulative Freq.(\%)\\
\hline
1 - 2 & 14 & 1.50   \\
3 - 4 & 22 & 2.40  \\
5 - 6 & 104 & 11.1  \\
7 - 8 & 422 & 45.1  \\
9 - 9.9 & 374 & 40.0 \\
\hline
\end{tabular}
\end{table}

Table 2 highlights the Mean Time $M.Time$ required for provenance encapsulation for different sized files and different counts of vm-instances. The $M.Time$ is shown to be less than 10 seconds for the large number of vm-instances of 936. As identified in \cite{park2011ramp}, the benchmark of tolerable time for provenance encapsulation is 10 seconds for a large system. Taking that value as a benchmark, implementation of ProvCapsule and ProvOCal algorithms show that the time overhead is below 10 seconds for over 900 instances. The overhead is found to be 8.198 seconds on average which is acceptable. The standard deviation is found to be 1.434 which is desirable.

\begin{figure}[ht!]
\centering
\includegraphics[width=3.76in,natwidth=210,natheight=942]{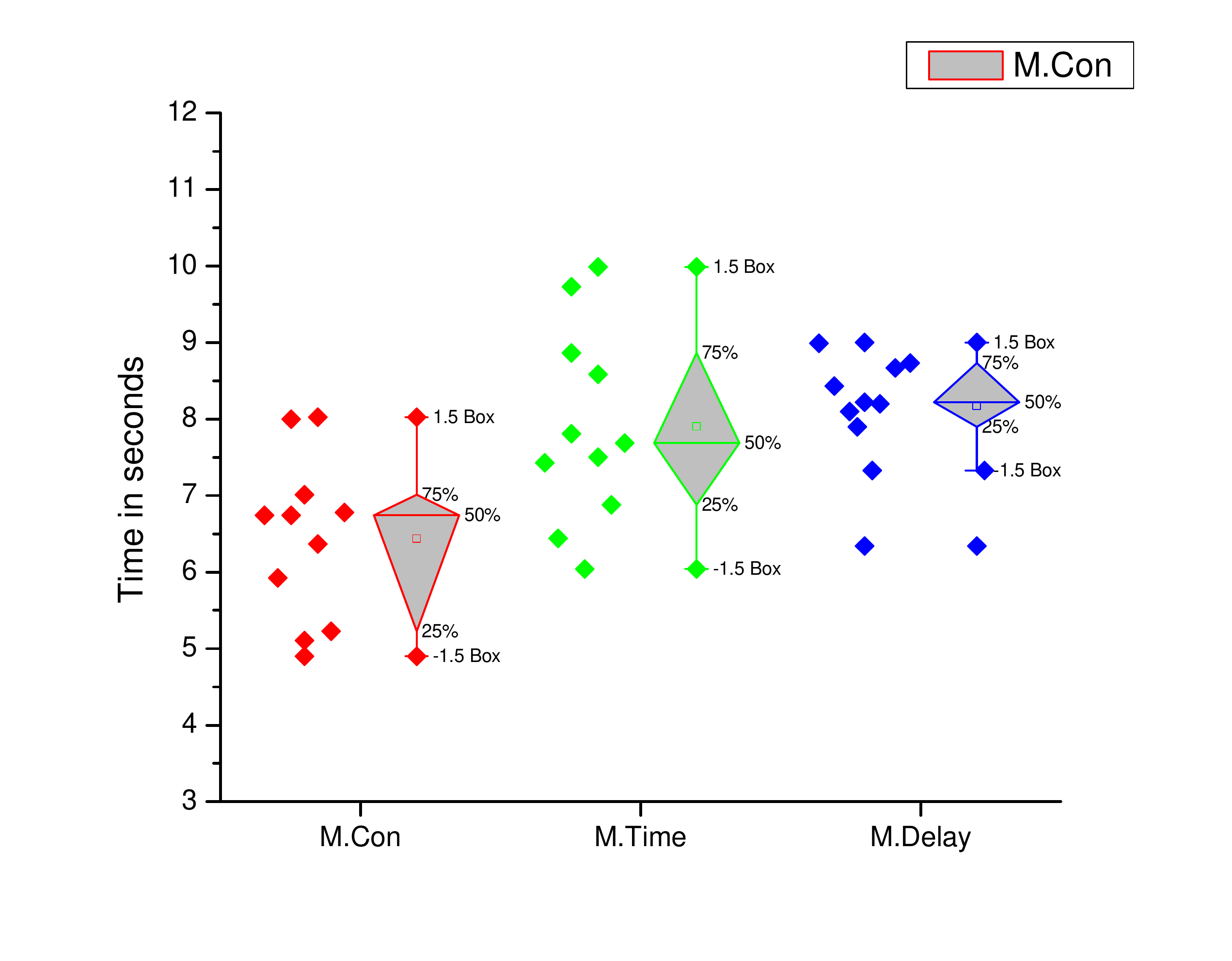}
\caption{Distribution of overhead}
\label{result6}
\end{figure}

An important aspect of provenance management regarding data leakage is the causality of information. Information from one process can be causally related to information of another process, hence the atomic actions can be causally combined to evaluate against pre-specified benchmarks and reduce false-positives.

Finally Table 3 shows the delay range and frequency of delay for all the vm-instances. It is seen that 45\% of the instances face an average delay of 7-8 seconds. The delay between 9-9.9 seconds is faced by 40\% of the instances for encapsulating provenance metafile. Hence the average delay is below 10 seconds for both the algorithms.

The variance and the distribution of the vm-instance classes are shown in Figure \ref{result6}. The variance is maximum for $M.Con$ whereas it is minimum for $M.Delay$ since the resources allocated in terms of memory and storage for each instance class was different. Hence the benchmark stated in \cite{park2011ramp} is satisfied.

\section{Conclusion}
\label{S7} \vspace{-4pt}

The paper aimed to ascertain a solution to the bottleneck of capturing provenance data in a widely decentralized system such as cloud computing. Algorithms for detecting provenance and encapsulating those on the specific objects have been proposed in this paper. Compared to traditional provenance detection techniques which function in standalone systems, the proposed active-threaded algorithms are capable of detecting provenance in a virtualized environment.

The performance analysis of the proposed algorithms show that the overhead incurred for capturing provenance using the proposed method is significantly low at 8.198 seconds with retries of 7, 3 and 7 respectively for 936 transactions. The standard deviation of 1.434 and cumulative frequency of 45.1\% show desirable performance of the algorithms.

As stated earlier, the proposed mechanism detects provenance for enabling forensic experts to use it in digital forensic investigation. Ensuring performance of the algorithms for preserving scalability of the cloud is a topic of future research interest.

\end{document}